\documentclass[journal]{IEEEtran}

\ifCLASSINFOpdf
\else
   \usepackage[dvips]{graphicx}
\fi

\usepackage{graphicx}
\usepackage[caption=false,font=footnotesize]{subfig}
\usepackage{subcaption}
\usepackage{amsmath,amssymb,amsfonts}
\usepackage{algorithmic}
\usepackage{epstopdf}
\usepackage{hyperref}
\usepackage{booktabs}
\usepackage{xcolor}

\begin{document}


\title{Cramer-Rao Bounds for Target Parameter Estimation in a Bi-Static IRS-Assisted Radar Configuration}

\author{S. S. Reddy, \IEEEmembership{Graduate Student Member, IEEE}, V. V. Reddy, \IEEEmembership{Member, IEEE.}

\thanks{S. Sanjeeva Reddy and Vinod Veera Reddy are affiliated with the Dept. of Electronics and Communication Engineering, International Institute of Information Technology, Bengaluru, India, 560100. \\
e-mail: sanjeevareddy.s414@ieee.org and vinod.reddy@iiitb.ac.in}}

\markboth{}
{Sanjeeva \MakeLowercase{\textit{et al.}}: Cramer-Rao Bounds for Target Parameter Estimation}
\maketitle


\begin{abstract}
The use of Intelligent Reflective Surfaces (IRS) to assist communication and sensing has proven cost-effective in challenging scenarios. For sensing, IRS is shown to sense non-line-of-sight (NLOS) and stealth targets, albeit with significant loss due to the four-hop path model. Amongst the available IRS-assisted configurations, we consider a three-hop model in which the IRS redirects the scattered target response towards the mono-static radar. With the IRS spatially displaced from the radar, this configuration mimics a bi-static radar. While target detection has been studied in this configuration, parameter estimation has not been investigated to date. To this end, we first develop the signal model for this configuration and derive the Cramer Rao Bounds (CRB) for target parameters. The dependence of CRB on system parameters such as SNR, number of snapshots, number of IRS elements and their weights is brought forward through extensive simulations. This study can enable a designer to customize the system parameters to meet the requirements. It also serves as a benchmark for parameter estimation techniques developed for this configuration.
\end{abstract}




\section{Introduction}
Radar has historically been a vital sensing technology capable of \textit{active} and \textit{passive} detection, target tracking, high-resolution imaging, and, increasingly, sensing in \textit{Non-Line-of-Sight} (NLoS) environments \cite{passive,isarimaging,activepassive,NLos}. Conventional mono-static radars usually perform well when targets are within their Line-of-Sight (LoS), enabling precise measurement of range, Doppler shifts, and directions. However, when obstacles block the LoS or targets exhibit stealth behaviour, the effectiveness of traditional radar diminishes significantly or results in complete detection failure, driving the innovation of new sensing architectures.

A common method for enabling NLoS radar sensing involves the use of an \textit{Intelligent Reflective Surface} (IRS). An IRS is made up of passive, adaptable reflecting elements that can steer incident electromagnetic waves, creating controllable pathways beyond the LoS when positioned appropriately. This \textit{IRS-assisted radar} technology is well suited in defence, automotive sensing, and modern wireless systems to expand the field-of-view (FOV) and improve NLoS target detection capabilities~\cite{defence,automotive,comm}. Meanwhile, the presence of an IRS introduces multiple possible signal paths from the target to the radar. Specifically, deploying an IRS results in four possible propagation paths~\cite{crbangle}:

\begin{align*}
\text{(P1)}~& \text{Radar} \rightarrow \text{Target} \rightarrow \text{Radar},\\
\text{(P2)}~& \text{Radar} \rightarrow \text{IRS} \rightarrow \text{Target} \rightarrow \text{IRS} \rightarrow \text{Radar}, \\
\text{(P3)}~& \text{Radar} \rightarrow \text{Target} \rightarrow \text{IRS} \rightarrow \text{Radar}, \\
\text{(P4)}~& \text{Radar} \rightarrow \text{IRS} \rightarrow \text{Target} \rightarrow \text{Radar}.
\end{align*}

Most IRS-assisted radar studies typically focus on the dominant LoS component (P1) and the IRS-enabled NLoS component (P2)~\cite{p1p2}. NLoS sensing via (P2) has been widely studied, including the derivation of Cramér--Rao Lower Bounds (CRLB) for selected target parameters~\cite{p2crb}. Despite the beamforming gain yielded by IRS weights in this configuration, it cannot fundamentally mitigate the severe path loss inherent to the four-hop channel, due to the compounded attenuation over the radar--IRS and IRS--target links~\cite{p2attenuation}. This configuration is therefore not suitable for sensing low radar cross-section (RCS) targets and stealth targets.

To mitigate performance degradation due to path loss and scattering from stealth targets, bi-static and multi-static radar configurations are considered. Unlike mono-static radar, in which the transmitter (Tx) and receiver (Rx) are co-located, bi-static and multi-static radars spatially separate the Tx and Rx (s), thereby forming a distributed sensing architecture~\cite{bistatic,multistatic,bimultistatic}. This separation enables the receiver to observe the target from a viewpoint that is different from the LoS path between the transmitter and target. In multi-static reception, multiple spatially separated receivers observe the target returns, which can improve detection probability and tracking performance due to spatial diversity~\cite{multistatictracking1,multistatictracking2}. Such configurations are particularly useful for detecting low-observable (stealth) targets and for sensing space objects, making them attractive for modern defence and space applications~\cite{bistaticuav,bistaticspace}. Despite these advantages, they introduce important practical limitations, most notably stringent time and phase synchronization requirements between the Tx and distributed Rx (s), along with increased hardware cost and maintenance overhead compared to the mono-static radar.

Upon investigating these trade-offs, we observe the need for a configuration that (i) reduces the effective path loss relative to the double-IRS (P2) route, and (ii) avoids inter-node time synchronization errors inherent in bi-static/multi-static radars. In this work, we focus on the \emph{third path} (P3), i.e., \textbf{Radar $\rightarrow$ Target $\rightarrow$ IRS $\rightarrow$ Radar}, which is often overlooked but is extremely useful. As will be shown, this path is analogous to a bi-static/multi-static observation geometry as the IRS provides a spatially displaced ``listening'' viewpoint relative to the radar, yet it remains a passive structure. This motivates a bi-static-like configuration based on IRS-assisted radar. This configuration has been recently presented in~\cite{danilopaper} for stealth target detection. However, target-parameter estimation within this framework (P3) has not been studied in the literature.

In this work, the IRS-assisted radar configuration is inspired by the bi-static radar configuration. The position of the IRS panel will coincide with the position of the bi-static receiver which is spatially away from the mono-static radar transceiver. This position is determined by the desired FOV, which is the intersection of the transmitter and the IRS beam patterns. In addition to this requirement, this configuration requires a line-of-sight path between the IRS and the radar transceiver for the IRS to redirect the target's response towards the radar. A dedicated receive chain can process the signal received from this channel. The sensing chain therefore leverages (P3), mimicking certain advantages of bi-static/multi-static radar while maintaining a mono-static radar front-end, thereby eliminating explicit time synchronization errors between Tx and Rx. Moreover, employing an IRS panel to realize a bi-static-like sensing geometry in challenging scenarios (e.g., stealth target detection) introduces a novel paradigm: from a geometric sensing standpoint, the IRS-assisted propagation path can be interpreted as effectively \emph{translating the sensing aperture} to the IRS location and subsequently mapping the estimated target coordinates back into the original radar reference frame.

The contributions of the paper are summarized as follows:
\begin{itemize}
    \item The IRS-assisted radar configuration is presented in comparison to bi-static/multi-static radar and existing IRS-assisted radars. The associated signal model is systematically developed and reformulated into a parametrized model suitable for the Cramér-Rao Bound(CRB) derivation.
    \item We derive CRB for the target angle parameters (azimuth and elevation) under the proposed signal model.
    \item Extensive simulation studies are performed to validate the performance of the IRS-assisted radar configuration under different geometric and channel conditions.
\end{itemize}

The paper is organized as follows: Section~\ref{radarsetup} explains in detail the identified IRS-assisted radar configuration. Section~\ref{signalmodel} provides the detailed mathematical derivation of the signal model and the target parameters. Based on these target parameters, the theoretical CRB limits are derived in Section~\ref{crlb}. Section~\ref{results} discusses the numerical results in relation to the theoretical bounds, and Section~\ref{conclusion} concludes the paper.


\section{IRS-Assisted Radar Setup}\label{radarsetup}
\begin{figure*}[!t]
    \centering
    \begin{minipage}[b]{0.3\textwidth}
        \centering
        \subfloat[Configurations: (i). Bi-static, (ii). Single IRS, (iii). Multi-static, (iv). Multiple IRS.\label{compmodel}]{
            \includegraphics[width=\linewidth]{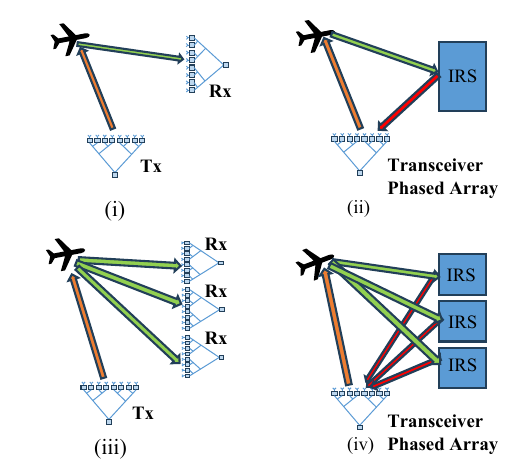}
        }
    \end{minipage}\hfill
    \begin{minipage}[b]{0.3\textwidth}
        \centering
        \subfloat[\label{system}]{
            \includegraphics[width=\linewidth]{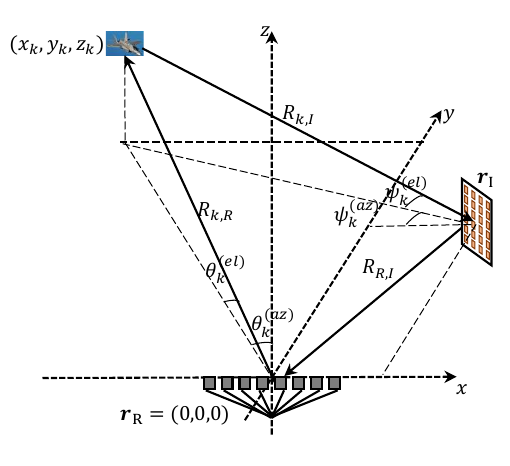}
        }
    \end{minipage}\hfill
    \begin{minipage}[b]{0.3\textwidth}
        \centering
        \subfloat[\label{Fig:BistaticFOV}]{
            \includegraphics[width=\linewidth]{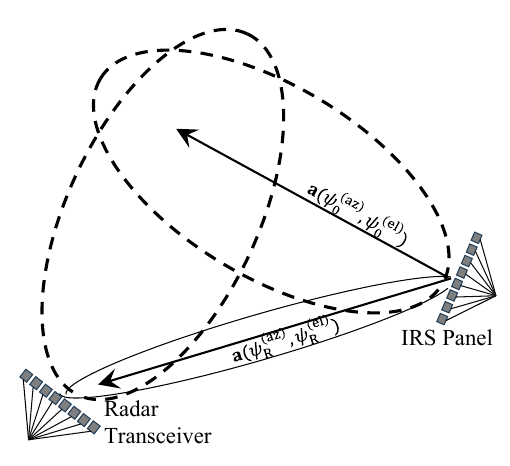}
        }
    \end{minipage}

    \caption{(a) Comparison of radar configurations, (b) IRS-static radar setup, and (c) IRS-static radar field-of-view (FOV).}
    \vspace{-0.5cm}
    \label{fig:combined_three_column}
\end{figure*}

Bi-static radar separates the transmitter and receiver, forming a signal path with the target as shown in Fig.~\ref{fig:combined_three_column}(a)(i), unlike mono-static radar’s setup . This concept generalizes to multi-static radar in Fig.~\ref{fig:combined_three_column}(a)(iii), which uses multiple spatially dispersed transmitters and receivers to gain spatial diversity. Multi-static systems observe targets from several angles, enhancing the detection of complex or stealth targets whose Radar Cross Section (RCS) varies with aspect angle. However, effectively exploiting these advantages requires careful spatial arrangement of the receivers, as discussed in \cite{Richards2005Book}.

The analogous IRS-assisted radar setup replaces traditional bi-/multi-static receiver nodes with low-cost IRS panels, as shown in Fig.~\ref{fig:combined_three_column}(a)(ii) and  Fig.~\ref{fig:combined_three_column}(a)(iv), which redirect the scattered signals toward nearby radar transceivers~\cite {danilopaper}. By carefully designing the IRS element weights, the system achieves sufficient beamforming gain to offset longer propagation paths. As shown in Section V, the parameter estimation performance depends on various factors such as IRS design and placement, target placement and characteristics. Fixed IRS placement simplifies radar-IRS channel estimation, and facilitates circumventing synchronization issues typically observed in conventional bi-/multi-static radars.

Amongst the four IRS-assisted configurations discussed in~\cite{irsmimoradar1} ((P1)-(P4)), the considered IRS-assisted configuration caters to (P1) and (P3) signal paths. This enables the detection of both conventional and stealth targets. Unlike prior IRS-assisted radar work~\cite{irsmimoradar2}, which places the IRS near the radar or target to boost signal strength and relies on a 4-path model (P2)~\cite{p2crb}, the proposed design positions the IRS panel away from the radar and operates it in simplex mode so that the target response follows the Target → IRS → Radar path. This yields an effective three-path model (P3), reduces path loss compared to conventional four-path IRS-assisted systems, and functionally emulates a bi-static radar configuration. We therefore address this configuration of interest as ``\textit{IRS-static radar}" in the rest of the paper.

\section{Signal Model} \label{signalmodel}
Consider a scenario with $K$ targets, with the radar located at $\mathrm{\textbf{r}}_{\mathrm{R}} = [0, 0, 0]^T$ of the coordinate system. Let the $k$th target parameters-namely range, azimuth angle, and elevation angle - be given by $\textbf{r}_k$, $\theta^{(\mathrm{az})}_k$, and $\theta^{(\mathrm{el})}_k$, respectively. Without loss of generality, we consider a 2-D IRS panel, with $L=L_hL_v$ elements, where $L_h$ and $L_v$ denote the number of elements along the horizontal and vertical dimensions, respectively, located at $\mathbf{r}_{\mathrm{I}}$ in the radar coordinate system. Let the $k$th target subtend azimuth and elevation angles $\psi^{(\mathrm{az})}_k$ and $\psi^{(\mathrm{el})}_k$, respectively, as seen from the IRS coordinate system, illustrated in Fig.~\ref{fig:combined_three_column}(b). We assume a phased-array radar with $M_T = M_R$ transceivers co-located at $\mathbf{r}_{\mathrm{R}}$.

Consider $\mathbf{s}(t) =  \exp\!\left[j\left(2\pi f_{0}t+\pi\mu t^{2}\right)\right] \operatorname{rect}\!\left(\frac{t-T_p/2}{T_p}\right). \in\mathbb{C}^{M_T\times 1}$ as the beamformed transmit signal, where $f_0,\mu, T_p$ are the operating frequency, frequency slope and pulse width of the signal, respectively. This signal, beamformed towards the desired direction (FOV), interacts with the $k$th target. We focus on two scattered components of interest, the monostatic return and the component propagating towards the IRS panel from the target (radar-to-target-to-IRS). Denoting the channel between the radar and the $k$th target by $\mathbf{g}_k \in \mathbb{C}^{1 \times M_T}$, the monostatic (ms) component can be expressed as
\begin{equation}
\mathbf{y}^{\mathrm{(ms)}}(t,t_l) = \sum_{k=1}^K \alpha^{\mathrm{(ms)}}_k \mathbf{g}_k\mathbf{g}_k^{H}\mathbf{s}(t-\tau'_{k,l}),
\end{equation}
where ${}^H$ denotes the conjugate transpose and the channel $\mathbf{g}_k$ is assumed reciprocal. Here, $\alpha^{\mathrm{ms}}_k$ denotes the monostatic target response strength, $t$ and $t_l$ represent the fast-time and slow-time variables, respectively, and $\tau'_{k,l} = 2(\|\mathbf{r}_k-\mathbf{r}_{\mathrm{R}}\|_2)/c$ is the two-way path delay between the radar and the $k$th target located at $\mathbf{r}_k$.

Next, we define the channel between the $k$th target and the IRS panel as $\mathbf{h}_k\in\mathbb{C}^{L_hL_v\times 1}$, and the channel between the IRS panel and the phased-array transceiver as $\mathbf{F}\in \mathbb{C}^{M_R\times L}$. The signal received at the radar via the IRS path (denoted as ``is'') is given by
\begin{equation}
\mathbf{y}^{\mathrm{is}}(t,t_l) = \sum_{k=1}^K \alpha^{\mathrm{(is)}}_k \mathbf{F}\mathbf{D}\mathbf{h}_k{\mathbf{g}_k}^{H}\mathbf{s}(t-\tau_{k,l}).
\end{equation}
Let $\alpha^{\mathrm{is}}_k$ denote the response strength associated 
with the IRS-reflected path and the diagonal matrix
\(\mathbf{D} = \mathrm{diag}(d_1, \dots, d_l, \dots, d_{L_hL_v}) 
\in \mathbb{C}^{L_hL_v \times L_hL_v}\) contains the reflection coefficients of all IRS elements. Each coefficient is expressed as
\(d_l = \beta_l e^{-j\rho_l},\) where $|\beta_l| = 1$ denotes the unit-modulus amplitude, and $\rho_l$ is the phase shift induced by the $l$-th reflecting element of the IRS, for all \(1 \le l \le L_h L_v\). It is important to note that the total path delay
\begin{equation}
\tau_{k,l} = \frac{\|\mathbf{r}_k-\mathbf{r_{\mathrm{R}}}\|_2+\|\mathbf{r}_k-\mathbf{r_{\mathrm{I}}}\|_2+\|\mathbf{r_{\mathrm{R}}}-\mathbf{r_{\mathrm{I}}}\|_2}{c},
\label{dist delay}
\end{equation}
is greater than  $\tau'_{k,l}$. Accordingly, the overall received signal at the transceiver can be written as
\begin{equation}
\mathbf{y}(t,t_l) = \mathbf{y}^{\mathrm{(ms)}}(t,t_l) + \mathbf{y}^{\mathrm{(is)}}(t,t_l) + \mathbf{v}(t,t_l),
\end{equation}
where $\mathbf{v}(t,t_l)$ is the Additive White Gaussian noise (AWGN) vector. 

For conventional targets, $\alpha^{\mathrm{(ms)}}_k$ is typically large; hence, returns from both paths must, in principle, be processed. However, because the path delays differ between the monostatic path and the path via the IRS, the signals must be processed independently. The separation of the monostatic component from the IRS response can be accomplished by beam-space processing at the radar receiver or by having a dedicated receive array looking towards the IRS panel at the radar. Since parameter extraction from conventional targets using the monostatic path is well studied, we focus our study on the parameter estimation from the IRS path. The signal received by this array from only the IRS path is 
\begin{equation}
\small
\label{SignalIRSpath}
\begin{split}
\mathbf{\overline{y}}(nT_s,t_l) &= \sum_{k=1}^K \alpha^{\mathrm{(is)}}_k \mathbf{F}\mathbf{D}\mathbf{h}_k\mathbf{g}_k^{H}\mathbf{s}(nT_s-\tau_{k,l}) + \mathbf{v}(nT_s,t_l).
\end{split}
\end{equation}

It is observed that the IRS weights in $\mathbf{D}$ influence the received signal in  (\ref{SignalIRSpath}). We next investigate the design of the IRS coefficients and the reconstruction of the underlying signal model in a manner that facilitates estimation of the target parameters in the following subsections.

\subsection{Design of IRS Weights}\label{irsweightdesign}
To achieve performance analogous to a bi-static/multi-static radar, the IRS-static radar must incorporate the effective field-of-view (FOV) of the corresponding distributed configuration into its design. From the IRS-static radar depicted in Fig.~\ref{fig:combined_three_column}(c), we note that the bi-static receiver is replaced by the IRS panel. To realize the desired FOV, the bi-static radar receiver would be oriented toward the direction $\big(\psi^{\left(\mathrm{az}\right)}_0,\psi^{\left(\mathrm{el}\right)}_0\big)$. For the IRS-static radar, the IRS weights must be configured such that signals received from this angular region are redirected toward the radar receiver, which is located at an angle $\big(\psi^{\left(\mathrm{az}\right)}_{\mathrm{R}},\psi^{\left(\mathrm{el}\right)}_{\mathrm{R}}\big)$ relative to the IRS panel. The coherent signal arrival at the receive array serves as the design objective for IRS weights. Defining the diagonal matrices $\mathbf{A}_0 = \mathrm{diag}\{\mathbf{a}(\psi^{\left(\mathrm{az}\right)}_0,\psi^{\left(\mathrm{el}\right)}_0)\}$ and $\mathbf{A}_{\mathrm{R}} = \mathrm{diag}\{\mathbf{a}(\psi^{\left(\mathrm{az}\right)}_{\mathrm{R}},\psi^{\left(\mathrm{el}\right)}_{\mathrm{R}})\}$, the diagonal IRS weight matrix $\mathbf{D}$ should satisfy
\begin{equation}
    \mathbf{A}_0 \mathbf{D}\mathbf{A}_{\mathrm{R}} = \mathbf{I},
\end{equation}
where $\mathbf{I}$ is the identity matrix. The resulting solution is
\begin{equation}
\label{WeightsSoln}
    \mathbf{D}=\mathbf{A}_0^*\mathbf{A}_{\mathrm{R}}^*,
\end{equation}
where `*' denotes complex conjugation. For targets in the vicinity of $\big(\psi^{\left(\mathrm{az}\right)}_0,\psi^{\left(\mathrm{el}\right)}_0\big)$, the IRS panel redirects the response back to the radar with some beamforming gain, attaining maximum gain when the target is located exactly in the direction $\big(\psi^{\left(\mathrm{az}\right)}_0,\psi^{\left(\mathrm{el}\right)}_0\big)$. Although these weights are derived from the steering vectors, they exhibit a beam pattern similar to that of a conventional beamformer. The designed IRS weights have continuous phase, whereas current state-of-the-art IRS panels permit only quantized phase levels. The effect of this phase quantisation on the CRB is studied in Section.~\ref{results}.

\subsection{Reconstruction of Signal Model}\label{reconstruction}
Stacking the received signal in (\ref{SignalIRSpath}) over $C$ chirps, we obtain the radar data cube $\mathbf{\overline{Y}}\in\mathbb{C}^{M\times N \times C}$, where $N$ denotes the number of fast-time samples. Performing the discrete Fourier transform (DFT) along the slow-time dimension, we to obtain $M_R$ Range-Doppler (RD) maps,
$\mathbf{\overline{Y}}_{\mathrm{RD}} = \mathcal{F}_C\{\mathbf{\overline{Y}}\}$, where $\mathcal{F}_C\{ . \}$ is a $C$-point DFT operator. Due to the superposition of the $K$ target responses, $\mathbf{\overline{Y}}_{\mathrm{RD}} = \sum\limits_{k=1}^{K}\mathbf{\overline{Y}}^k_{\mathrm{RD}}$, where $\mathbf{\overline{Y}}^k_{\mathrm{RD}}$ denotes the RD contribution due to the $k$th target alone. Using either constant false-alarm-rate (CFAR) detector (or its variants)~\cite{Richards2005Book} or technique developed in~\cite{danilopaper}, one can obtain $K_P\leq K$ RD bins corresponding to true detections; with some bins potentially containing multiple targets. Therefore, the total number of targets $K = \sum_{k=1}^{K_P}P_k$, where $P_k$ denotes the number of targets present in the $k$th detected RD bin, $(\hat{R}_k,\hat{v}_k)$. Since we perform angle estimation at each of these $K_p$ range bins separately, range-angle coupling does not exist. Note that $\hat{R}_k=\frac{c\tau_{k,l}|_{l=0}}{2}$ corresponds to the round-trip distance via the IRS, while $\hat{v}_k$ denotes the relative bi-static velocity of the target with respect to the radar and the IRS.

The observation vector for the $k$th RD bin across the $M_R$ receivers is given by
\begin{equation}
\label{SparseObsVec1}
\mathbf{\overline{y}}_k = \mathbf{FD}\sum_{p=1}^{P_k}\alpha_p^{(is)}\mathbf{h}_p \mathbf{g}_p^{H}\mathbf{s} + \mathbf{\bar{v}}_k.
\end{equation}
We denote the LoS channel between the radar and the $p$th target as $\mathbf{g}_p = \alpha_p^{(g)}\mathbf{a}_{M_T}(\theta^{(\mathrm{az})}_p,\theta^{(\mathrm{el})}_p)$, where $\mathbf{a}_{M_T}(\theta^{(\mathrm{az})}_p,\theta^{(\mathrm{el})}_p)$ is the steering vector toward the target as seen from the transmit array. Similarly, we write $\mathbf{h}_p = \alpha_p^{(h)}\mathbf{a}_{L_hL_v}\left(\psi^{(\mathrm{az})}_p,\psi^{(\mathrm{el})}_p\right)$, where $\mathbf{a}_{L_hL_v}\left(\psi^{(\mathrm{az})},\psi^{(\mathrm{el})}\right) = \mathbf{a}_{L_v}\left(\psi^{(\mathrm{az})}\right) \otimes \mathbf{a}_{L_h}\left(\psi^{(\mathrm{el})}\right)$ is the steering vector toward the target as observed from the IRS panel, and $\otimes$ denotes the Kronecker product. $\alpha_p^{(is)}$ is the target scattering response through the IRS path (proportional to bi-static RCS). The effect of $\alpha_p^{(is)}$ in the received signal is directly related to the SNR as per the bi-static radar range equation. In this work, we therefore study the parameter estimation performance against SNR which in turn reflects the performance against the RCS. Denoting $\alpha_p = \alpha_p^{(h)}\alpha_p^{(g)}$ as the combined path loss from radar-to-target-to-IRS and accounting for the IRS-to-Radar loss within the channel matrix (\textbf{F}), (\ref{SparseObsVec1}) can be expressed as
\begin{multline}
\mathbf{\overline{y}}_k = \mathbf{FD}\sum_{p=1}^{P_k} \alpha_p^{(is)}\alpha_p\mathbf{a}_{L_hL_v}\left(\psi_p^{\left(\mathrm{az}\right)},\psi^{\left(\mathrm{el}\right)}_p\right) \\ \times \mathbf{a}_{M_T}\left(\theta^{\left(\mathrm{az}\right)}_p,\theta^{\left(\mathrm{el}\right)}_p\right)^H\mathbf{\overline{s}} + \mathbf{\bar{v}}_k.
\label{sig model}
\end{multline}  

With reference to Fig.~\ref{fig:combined_three_column}(b) for the detection obtained at ($\hat{R}_k$,$\hat{v}_k$), the parameters $\theta^{\left(\mathrm{az}\right)}_k,\theta^{\left(\mathrm{el}\right)}_k, \psi^{\left(\mathrm{az}\right)}_k,\psi^{\left(\mathrm{el}\right)}_k, \forall \ k \in [1,K]$ are to be estimated. Importantly, the 3-D target position can be determined using only a subset of these parameters. Let $R_{k,\mathrm{R}}$ and $R_{k,\mathrm{I}}$ denote the distances from the $k$th target to the radar and to the IRS, respectively. Then the estimated range inferred from the path delay satisfies
\begin{equation}
\hat{R}_k = R_{k,\mathrm{R}}+R_{k,\mathrm{I}}+R_{\mathrm{I},\mathrm{R}},
\end{equation}
where the distance between IRS and the radar $R_{\mathrm{I},\mathrm{R}}$ is assumed known. It follows that estimating $\psi^{(\mathrm{az})}_{k}$ and $\psi^{(\mathrm{el})}_{k}$, together with $R_{k,\mathrm{I}}$, yields the $k$th target position $\mathbf{r}_k' = [x_k', y_k',z_k']^T$ with respect to the IRS coordinate system. The target position in the radar coordinate system can then be obtained via
\begin{equation}
\label{TgtPosnWRTradar}
\mathbf{r}_k = \mathbf{r}_k'+\mathbf{r}_{\mathrm{I}}.
\end{equation}
Alternatively, the target position can be obtained directly by estimating $\theta^{\left(az\right)}_{k}$ and $\theta^{\left(el\right)}_{k}$ along with $R_{k,\mathrm{R}}$ from $\hat{R}_k$. 

It is evident from (\ref{sig model}) that the received signal corresponds to a point in 3-D space that is defined by the intersection of $\mathbf{a}_{L_vL_h}\left(\psi_p^{\left(\mathrm{az}\right)},\psi_p^{\left(\mathrm{el}\right)}\right)$ and $\mathbf{a}_{M_T}\left(\theta_p^{\left(\mathrm{az}\right)}\right)$, further constrained by the round-trip delay. (\ref{sig model}) can be compactly represented as:
\begin{equation}
    \overline{\textbf{y}}_k = \textbf{F}\textbf{D}\textbf{A}\overline{\textbf{x}} + \overline{\textbf{v}}_k,
    \label{compact sig model}
\end{equation}
where \(\overline{\textbf{x}} = \sum^{P_k}_{p=1}\alpha_p^{(is)}\alpha_p \textbf{a}_{MT}(\theta^{az}_p,\theta^{el}_p) \textbf{s}\) and the dictionary matrix $\mathbf{A} \in \mathbb{C}^{L \times G_R}$ corresponds to the angle-of-arrival space around the IRS such that
\begin{equation}
    \mathbf{A} = \big[\mathbf{a}(\psi_1), \dots , \mathbf{a}(\psi_{G_R})\big],
\end{equation}
where the angular grid points $[\psi_1, ...,  \psi_{G_R}]$ are expressed as ${\Psi} = \{\psi_z : \psi_z  = \left(\psi_z^{(\mathrm{az})},\psi_z^{(\mathrm{el})}\right),   \psi_z^{(\mathrm{az})} \in \left[- \pi/2, \pi/2\right], \psi_z^{(\mathrm{el})} \in \left[0, \pi/2\right]\ 1\leq z\leq G_R\}$. Based on the target field of view with respect to the IRS, the angular region (and hence the number of grid points, $G_R$) can be reduced. In the following section, Cramer-Rao Bounds are derived for the signal model presented above.


\section{Cramér-Rao Bounds}\label{crlb}
The target parameters to be estimated  from all detected observation vectors $\overline{\mathbf{y}}_k$ are
\begin{equation}
\begin{aligned}
\boldsymbol{\zeta} &= [\zeta_1, \zeta_2, \ldots, \zeta_K], \\
\end{aligned}
\label{target_parameters}
\end{equation}
where, \(\boldsymbol{\zeta}_i = [\psi_i^{(\mathrm{az})},\, \psi_i^{(\mathrm{el})}]^{T}\)is the target parameters correponding to $k$-th target and \(\boldsymbol{\mathbf{\zeta}} \in \mathbb{R}^{2\times K}\) is the target parameters for all $K$ targets as seen from the IRS. Our objective is to characterize the fundamental error bound, i.e., the minimum achievable error variance, for estimating these angle parameters under the proposed signal model using the CRB framework.

Several approaches exist for deriving CRB depending on the assumptions and the structure of the signal model~\cite{vantrees2002optimum}. In other IRS-assisted radar configurations, CRB has been derived for signal waveform estimation in~\cite{p1p2} and for target parameter estimation in~\cite{crbangle}. In this work, we derive the CRB for target parameter estimation assuming a deterministic transmit signal waveform in the (P3) IRS-assisted configuration. We develop CRB for IRS-based $\zeta_i, i\in[1,K]$, which is equivalent to the study w.r.t. the radar because of the positional equivalence observed in~(\ref{TgtPosnWRTradar}).

For CRLB analysis, the proposed signal model in (\ref{compact sig model}) is rearranged as
\begin{equation}
    \overline{\textbf{y}}_k = \overline{\textbf{A}}\overline{\textbf{x}} + \overline{\textbf{v}},
    \label{crlb sig model}
\end{equation}
where \(\overline{\textbf{A}}=\textbf{F}\textbf{D}\textbf{A}\). This compact form of the signal model ensures that only the target parameters as seen from the IRS, i.e., $ (\psi^{\mathrm{(az)}},\psi^{\mathrm{(el)}})$, are explicit while the rest are embedded within $\overline{\mathbf{A}}$.

The Fisher Information Matrix (FIM) associated with the above signal model and the parameter vector of interest is given by
\begin{equation}
\small
\mathbf{J} =
\begin{bmatrix}
\label{jmat}
\mathbf{J}_{\psi^{(\mathrm{az})},\psi^{(\mathrm{az})}} &
\mathbf{J}_{\psi^{(\mathrm{az})},\psi^{(\mathrm{el})}} \\
\mathbf{J}_{\psi^{(\mathrm{el})},\psi^{(\mathrm{az})}} &
\mathbf{J}_{\psi^{(\mathrm{el})},\psi^{(\mathrm{el})}}
\end{bmatrix}.
\end{equation}

To obtain $\mathbf{J}$, we first define the vector derivative for each column of $\overline{\textbf{A}}$ with respect to target parameters as
\begin{equation}
    \mathbf{Z}_i = \left[ \frac{\partial \mathbf{\overline{a}}(\boldsymbol{\zeta}_i)}{\partial \psi_{i}^{(\mathrm{az})}} \quad \frac{\partial \mathbf{\overline{a}}(\boldsymbol{\zeta}_i)}{\partial \psi_{i}^{(\mathrm{el})}} \right] \quad \in \mathbb{C}^{M_R\times 2}, i = 1, \cdots, K,
        \label{SVd}
\end{equation}
Stacking these derivatives of all target parameters for $K$ targets, we get
\begin{equation}
    \mathbf{Z} = \left[ \mathbf{Z}_1 \quad \mathbf{Z}_2 \quad \cdots \quad \mathbf{Z}_K \right] \quad \in \mathbb{C}^{M_R\times 2K},
\end{equation}
Defining 
\(\textbf{H}=\textbf{Z}^H \textbf{P}_{\frac{1}{\overline{\textbf{A}}}}\textbf{Z},\), 
where \( \textbf{P}_{\frac{1}{\overline{\textbf{A}}}} = \textbf{I} - \overline{\textbf{A}}(\overline{\textbf{A}}^H\overline{\textbf{A}})^{-1}\overline{\textbf{A}}^H\) is the orthogonal projection matrix of  \(\overline{\textbf{A}}\), (\ref{jmat}) can be compactly expressed as~\cite{vantrees2002optimum}:
\begin{equation}
    \textbf{J = }\frac{2N}{\sigma_w^2}\,
\mathrm{Re}\!\left[
\mathbf{H} \odot \left( \mathbf{S}_{\mathbf{f}}^{T} \otimes \mathbf{1}_{2\times 2} \right)
\right],
\label{jeq}
\end{equation}
where \(\sigma_w^2\), \(N\), \(S_f\)  and  \(\mathbf{1}_{2 \times 2}\) denote the noise variance, the number of snapshots, the source signal powers and  \(2\times 2\) matrix of ones, respectively. \(\odot\) is the Hadamard product operator. The resulting CRB expression is given by 
\begin{equation}
    \textbf{C}_{CR} =  \textbf{J}^{-1} = 
    \begin{bmatrix}
        \mathbf{C}_{\psi^{\mathrm{(az)}},\psi^{\mathrm{(az)}}} & \mathbf{C}_{\psi^{\mathrm{(az)}},\psi^{\mathrm{(el)}}} \\
        \mathbf{C}_{\psi^{\mathrm{(el)}},\psi^{\mathrm{(az)}}}& \mathbf{C}_{\psi^{\mathrm{(el)}},\psi^{\mathrm{(el)}}}
    \end{bmatrix}.
\end{equation}
The $\mathbf{C}_{CR}$ contains the variance corresponding to the azimuth and elevation of each target. Consequently, the CRB error bounds in degrees for azimuth and elevation estimates are obtained as
\begin{equation}
CRB (\psi^{(\mathrm{az})}) = \sqrt{diag(\textbf{C}_{\psi^{(\mathrm{az})},\psi^{(\mathrm{az})}})} ,
\end{equation}
\begin{equation}
CRB(\psi^{(\mathrm{el})}) = \sqrt{diag(\textbf{C}^{\psi^{(\mathrm{el})},\psi^{(\mathrm{el})}})}.
\end{equation}

Although the derived CRB resembles the standard form of a linear signal model, the critical contribution of the IRS path to the performance of the IRS-static radar configuration is embedded within the CRB as mentioned in Section~\ref{reconstruction}. Besides the dependence of CRB on the snapshots ($N$) and the SNR \((\sigma^2_w, S_f)\), the matrix $\textbf{H}$ embeds the system setup in relation with target positions as seen from the IRS. Since the considered IRS-static configuration is new, we employ the derived CRB to analyse the sensitivity of this configuration across various target and system scenarios for the target parameter estimation in the subsequent section.


\section{Results}\label{results}

The performance of the proposed IRS-static radar system is evaluated through numerical simulations. The radar transmitter (Tx) and receiver (Rx) are co-located and employ uniform linear arrays, while a rectangular IRS is deployed to provide an indirect sensing path for target localization. Unless otherwise stated, the geometric configuration, array specifications, IRS characteristics, and signal parameters used throughout the simulations are summarized in Table~\ref{tab:sim_parameters}.

\begin{table}[!b]
\centering
\caption{Simulation Parameters}
\label{tab:sim_parameters}
\renewcommand{\arraystretch}{1.15}
\begin{tabular}{@{} l l @{} }
\toprule
\textbf{Parameter} & \textbf{Value} \\
\midrule
Carrier frequency, $f_c$ & 24~GHz \\
Radar Tx/Rx location & $(0,0,0)$~m \\
Number of Tx/Rx antennas, $M_T/M_R$ & 50 \\
Radar element spacing, $d_{\mathrm{radar}}$ & $\lambda/2 = 6.25$~mm \\
IRS location, $\mathbf{r}_I$ & $[-50,\;100,\;0]^T$~m \\
IRS dimensions & $20\times18$ elements \\
Number of IRS elements, $L$ & 360 \\
IRS element spacing & $\lambda/2$ \\
Target location & $(x_t,y_t,z_t)$~m \\
Number of snapshots, $N$ & 128 \\
Transmit power, $P_t$ & 100~dBm \\
Signal bandwidth, $B$ & 250~MHz \\
Pulse duration, $T_p$ & 1~$\mu$s \\
Sampling frequency, $f_s$ & 1~GHz \\
SNR range & $-15$ to $15$~dB \\
\bottomrule
\end{tabular}
\end{table}

The IRS elements are arranged in the $y$--$z$ plane with half-wavelength spacing along both axes. The FOV of the IRS-static radar is the intersection of the transmit beam and the IRS look direction, as shown in Fig.~\ref{fig:combined_three_column}(c). Accordingly, the IRS weights are computed using (\ref{WeightsSoln}) so that the FOV is steered to the point $(\Psi^{(\mathrm{az})}_{0},\Psi^{(\mathrm{el})}_{0})$, and the reflected signal is directed toward the radar receive array at $(\Psi^{(\mathrm{az})}_{\mathrm{R}},\Psi^{(\mathrm{el})}_{\mathrm{R}})$. Being agnostic to any parameter estimation technique, we study the CRB against various system parameters.

In this study, the signal-to-noise ratio (SNR) is defined at the radar
receiver. It is the ratio of the average received signal power per antenna
element per snapshot to the noise power. For each configuration, we first
generate the noise-free received signal
$\mathbf{\overline{Y}} \in \mathbb{C}^{M_R \times N}$ from the signal model
in~\eqref{sig model}. Here $T$ is the number of snapshots. $\mathbf{\overline{Y}}$ includes all propagation effects:
the radar-to-target, target-to-IRS and IRS-to-radar path losses, the IRS phase
configuration and the target reflectivity. Its average power is
\begin{equation}
    P_s = \frac{1}{M_R T}\,\big\lVert \mathbf{\overline{Y}} \big\rVert_F^2 .
    \label{eq:signal_power}
\end{equation}
For a prescribed SNR, the noise variance is then set to \(\sigma_n^2 = \frac{P_s}{\mathrm{SNR_{Linear}}} \) and the noisy observation is formed as
\begin{equation}
    \mathbf{\overline{Y}} = \mathbf{\overline{Y}} + \mathbf{V},
    \label{eq:noisy_observation}
\end{equation}
where the entries of $\mathbf{V}$ are independent and identically distributed
circularly symmetric complex Gaussian random variables,
$[\mathbf{V}]_{m,t} \sim \mathcal{CN}(0, \sigma_n^2)$. Because $\sigma_n^2$ is
set from the received signal power, every configuration is evaluated at the
same SNR at the receiver.

\begin{figure}[!tb] 

    \centering 
    \includegraphics[width=0.8\linewidth]{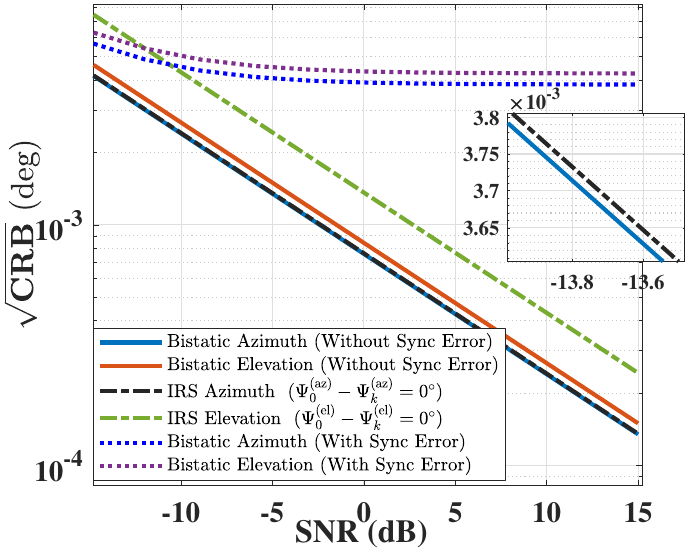}
    \caption{CRB versus SNR comparison for bi-static and IRS-static configurations.} 
    \vspace{-0.5cm} 
    \label{fig:rmse_bistatic} 
\end{figure}

Initially, we compare the performance of the proposed configuration with conventional bi-static radar whose transmitter and receiver mimic the geometry of the transmitter and IRS panel of the IRS-static configuration. To ensure a valid comparison, we consider bi-static radar with and without a time-synchronisation error of $ 1$ ps. We adopt the signal model and CRB derivation used for this comparison from~\cite{Yan2008}. As shown by the CRB curves in Fig.~\ref{fig:rmse_bistatic}, the performance of the IRS-static configuration is superior than the bi-static radar with time synchronization error while it approaches the CRB of the bi-static configuration without any synchronization error. In the IRS-static configuration, the azimuth angle estimation performance relies on the effective aperture that includes the IRS elements along the $y$-axis and the radar receiver elements along the $x$-axis. Despite additional path loss between the IRS and the radar, the CRB for azimuth is comparable with bi-static radar. However the CRB for elevation is marginally higher because the radar receiver is one-dimension. It is important to emphasize that a small time synchronization error in the bi-static radar can adversely affect parameter estimation performance. Moreover, as the SNR increases, the CRB curves exhibit saturation because the time-synchronization error introduces a model mismatch (bias) offset that cannot be mitigated by increasing signal power. Consequently, the estimation accuracy becomes synchronization-error limited, resulting in an error floor at high SNR.

\begin{figure*}[!t]
    \centering
    \begin{minipage}[b]{0.5\textwidth}
        \centering
        \subfloat[\label{fig:azi_error}]{
            \includegraphics[width=0.8\linewidth]{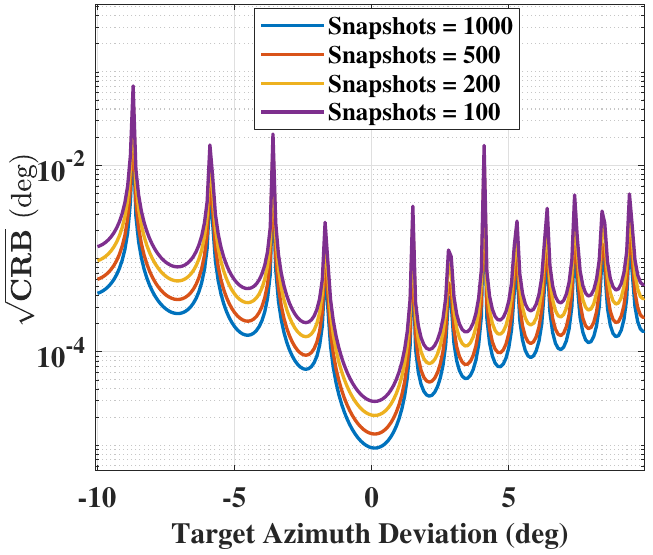}
        }
    \end{minipage}\hfill
    \begin{minipage}[b]{0.5\textwidth}
        \centering
        \subfloat[\label{fig:el_error}]{
            \includegraphics[width=0.8\linewidth]{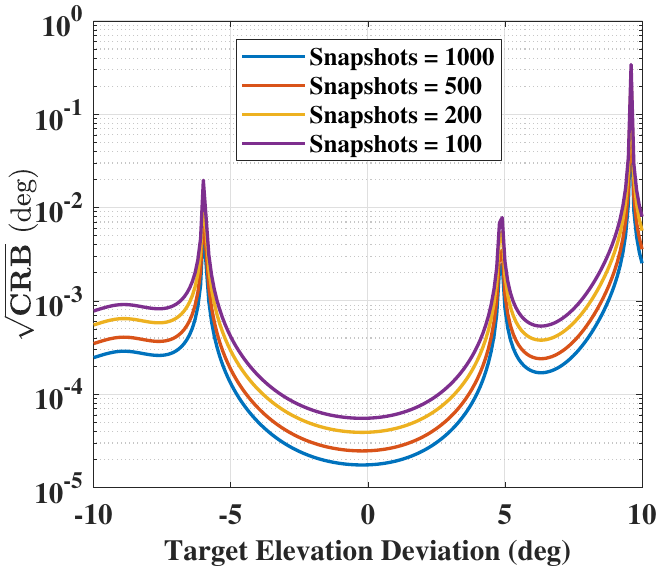}
        }
    \end{minipage}

    \caption{CRB versus angular mismatch between the target and IRS lookup direction in (a) Azimuth and (b) Elevation.}
    \label{fig:rmse_angle_mismatch}
\end{figure*}

Consider a scenario in which the target position coincides with the IRS look direction, i.e., ($\Psi^{(\mathrm{az})}_{0} = \Psi^{(\mathrm{az})}_{k}$, $\Psi^{(\mathrm{el})}_{0} = \Psi^{(\mathrm{el})}_{k}$). The variation in root-mean-squared error (CRB) when target azimuth and elevation deviates from the IRS look direction is studied in Fig.~\ref{fig:rmse_angle_mismatch} for a varying number of snapshots. It is observed that the CRB increases with increasing directional mismatch and exhibits an inverted pattern. The pre-steered IRS weights embedded within $\overline{\mathbf{A}}$ result in the Fisher information to vary in accordance with this pattern.

It is interesting to note that the CRB curve for azimuth angle deviation exhibits a narrow beam around $\Psi^{\mathrm{(az)}}_0$ as seen in Fig.~\ref{fig:rmse_angle_mismatch}(a), compared to that observed for elevation angle around $\Psi^{\mathrm{(el)}}_0$ in Fig.~\ref{fig:rmse_angle_mismatch}(b). This is because of the large effective aperture due to the IRS elements and the radar receive array in the x-y plane. For the elevation angle estimation, the wide CRB curve around $\Psi^{\mathrm{(el)}}_0$ in Fig.~\ref{fig:rmse_angle_mismatch}(b) can be attributed to the reduced effective aperture of IRS elements present along the z-plane. Furthermore, the inverted sidelobes do not exhibit significantly high CRB compared to the mainlobe, indicating target response from outside the FOV will receive a reasonable gain unless the target lies exactly on the beams’ nulls. From this observation, it can be inferred that the CRB performance, the FOV, and the sidelobe performance of the IRS-static radar can be altered through appropriate design of the IRS weights. Conversely, the required FOV along the azimuth and elevation angles have to be critical design parameters for IRS weights in this context. 

\begin{figure*}
    \centering
    \begin{minipage}[b]{0.5\textwidth}
        \centering
        \subfloat[\label{fig:rmse_snr}]{
            \includegraphics[width=0.8\linewidth]{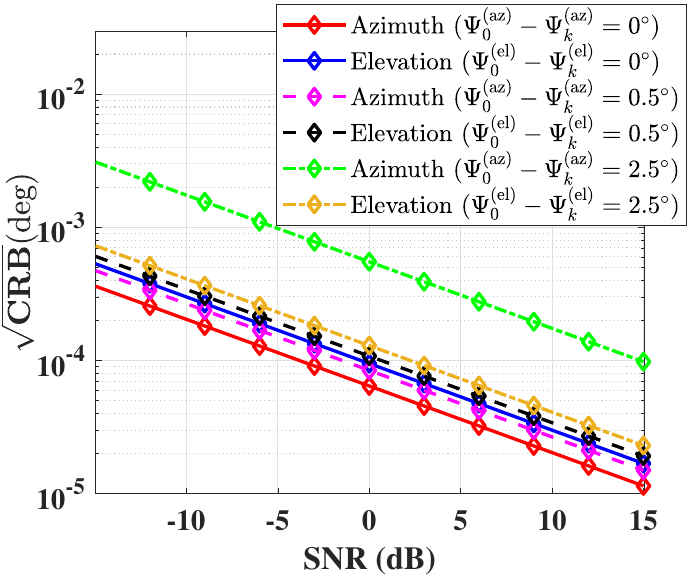}
        }
    \end{minipage}\hfill
    \begin{minipage}[b]{0.5\textwidth}
        \centering
        \subfloat[\label{fig:rmse_snapshots}]{
            \includegraphics[width=0.8\linewidth]{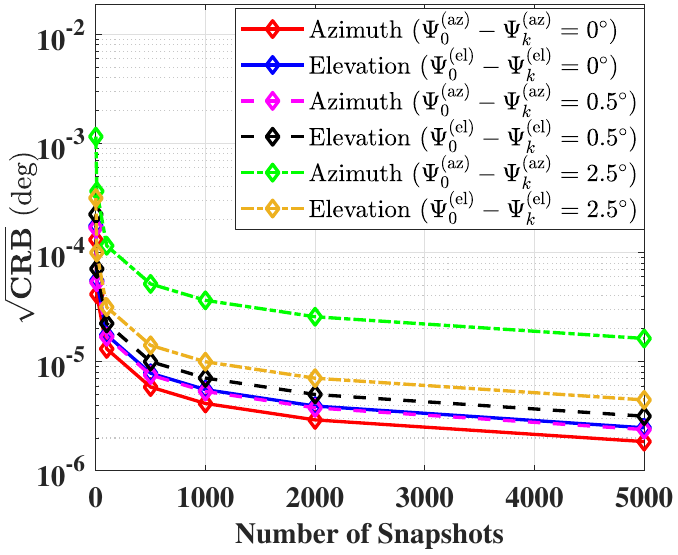}
        }
    \end{minipage}

    \caption{CRB versus (a) SNR and (b) Number of snapshots.}
    \label{fig:rmse_snr_ss}
\end{figure*}

We now study the CRB of azimuth and elevation angles of a single target as shown in Fig.~\ref{fig:rmse_snr_ss} against signal-to-noise ratio (SNR) varied from $-15~$dB to $15~$dB and number of snapshots varied from $1$ to $5000$. The CRB is plotted for a target position within the FOV ($\Psi^{(\mathrm{az})}_{0} = \Psi^{(\mathrm{az})}_{k}$ and $\Psi^{(\mathrm{el})}_{0} = \Psi^{(\mathrm{el})}_{k}$) and and two cases where the target is away from the FOV: (i)($\Psi^{(\mathrm{az})}_{0} - \Psi^{(\mathrm{az})}_{k}=0.5^\circ$,  and $\Psi^{(\mathrm{el})}_{0} - \Psi^{(\mathrm{el})}_{k}=0.5^\circ$) and (ii) ($\Psi^{(\mathrm{az})}_{0} - \Psi^{(\mathrm{az})}_{k}=2.5^\circ$,  and $\Psi^{(\mathrm{el})}_{0} - \Psi^{(\mathrm{el})}_{k}=2.5^\circ$).

A monotonous CRB decrease is observed in Fig.~\ref{fig:rmse_snr_ss}(a) with increasing SNR. For $\Psi_0=\Psi_k$, the CRB error for azimuth angle is distinctly lesser than the elevation angle over the entire SNR range due to the larger effective aperture along the x-y dimension over the z-dimension as discussed earlier. Similarly, Fig.~\ref{fig:rmse_snr_ss}(b) examines the dependence of azimuth and elevation CRB on the number of snapshots. Both error metrics decrease as the number of snapshots increases from the order of hundreds to the order of thousands, with progressively diminishing gains thereafter, once the covariance estimate becomes more stable.

When we specifically observe the plots corresponding to deviation of $0.5^\circ$, i.e., $\Psi^{(\mathrm{az})}_{0} - \Psi^{(\mathrm{az})}_{k}=0.5^\circ$ and $\Psi^{(\mathrm{el})}_{0} - \Psi^{(\mathrm{el})}_{k}=0.5^\circ$ (azimuth--pink and elevation--black), one can note that the CRB is higher than the condition when $\Psi^{(\mathrm{az})}_{0} = \Psi^{(\mathrm{az})}_{k}$ and $\Psi^{(\mathrm{el})}_{0} = \Psi^{(\mathrm{el})}_{k}$ because the target is off the IRS look direction. The azimuth CRB is lower than elevation CRB, similar to the previous case, as both these deviations lie in the main lobe corresponding to Fig.~\ref{fig:rmse_angle_mismatch}(a) and Fig.~\ref{fig:rmse_angle_mismatch}(b). For the case where $\Psi^{(\mathrm{az})}_{0} - \Psi^{(\mathrm{az})}_{k}=2.5^\circ$ and $\Psi^{(\mathrm{el})}_{0} - \Psi^{(\mathrm{el})}_{k}=2.5^\circ$, the corresponding plots (azimuth--green and elevation--gold) in Fig.~\ref{fig:rmse_snr_ss}(a) and Fig.~\ref{fig:rmse_snr_ss}(b) reveal that the CRB curves are higher than those of the $0^\circ$ and $0.5^\circ$ cases. This is due to the deviation from the look-up direction being larger compared to the $0.5^\circ$ case, resulting in larger loss in directional gain at the IRS. Furthermore, it is interesting to note that the elevation CRB is less than the azimuth CRB. This is because the target azimuth angle lies in the first sidelobe of the pattern observed in Fig.~\ref{fig:rmse_angle_mismatch}(a) while the target elevation angle still resides in the mainlobe pattern shown in Fig.~\ref{fig:rmse_angle_mismatch}(b).

\begin{figure}[!tb]
    \centering
    \includegraphics[width=0.8\linewidth]{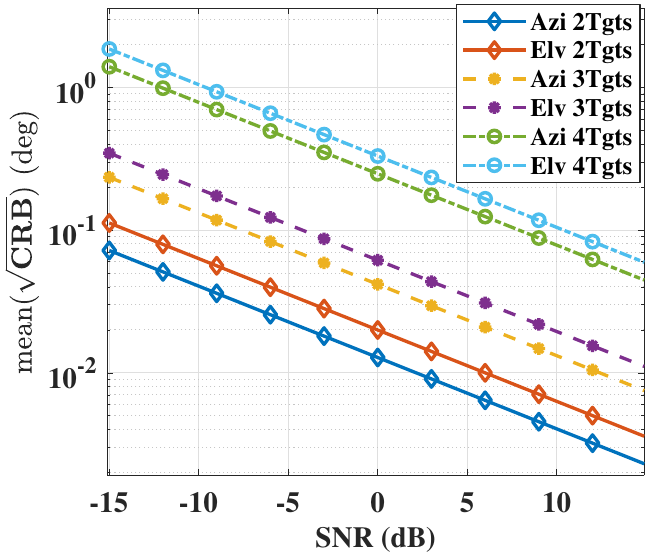}
    \caption{Average CRB versus SNR for different number of closely spaced targets.}
    \vspace{-0.5cm}
    \label{fig:rmse_ntgts}
\end{figure}

Next, we consider the presence of multiple targets within the field of view (FOV). We simulate scenarios containing 2, 3, and 4 closely-spaced targets, with the targets their response overlapping in adjacent range bins. The impact of multiple targets is significant only when they fall within the same or neighboring range bins, since angle estimation is performed on a per–range-bin basis after applying the detection stage. Consequently, as the number of closely-spaced targets increases, inter-target interference becomes more pronounced, and the estimation capability degrades, leading to larger CRB values, as shown in Fig.~\ref{fig:rmse_ntgts}.

\begin{figure*}[!ht]
    \centering
    \includegraphics[width=0.9\linewidth]{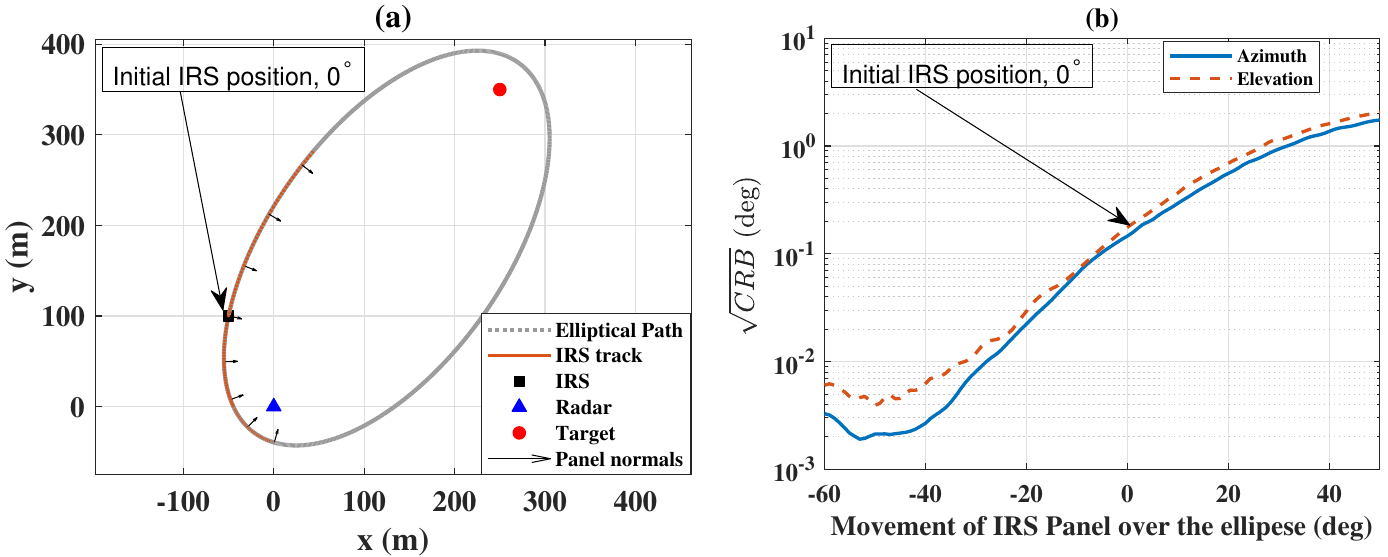}
    \caption{Study of CRB for IRS placement over the ellipse (a) Geometrical Representation of IRS Placement and (b) Performance of CRB as a function of IRS Placement.}
    \vspace{-0.5cm}
    \label{fig:rmse_irs_placement}
\end{figure*}

\begin{figure*}[h]
    \centering
    \includegraphics[width=0.9\linewidth]{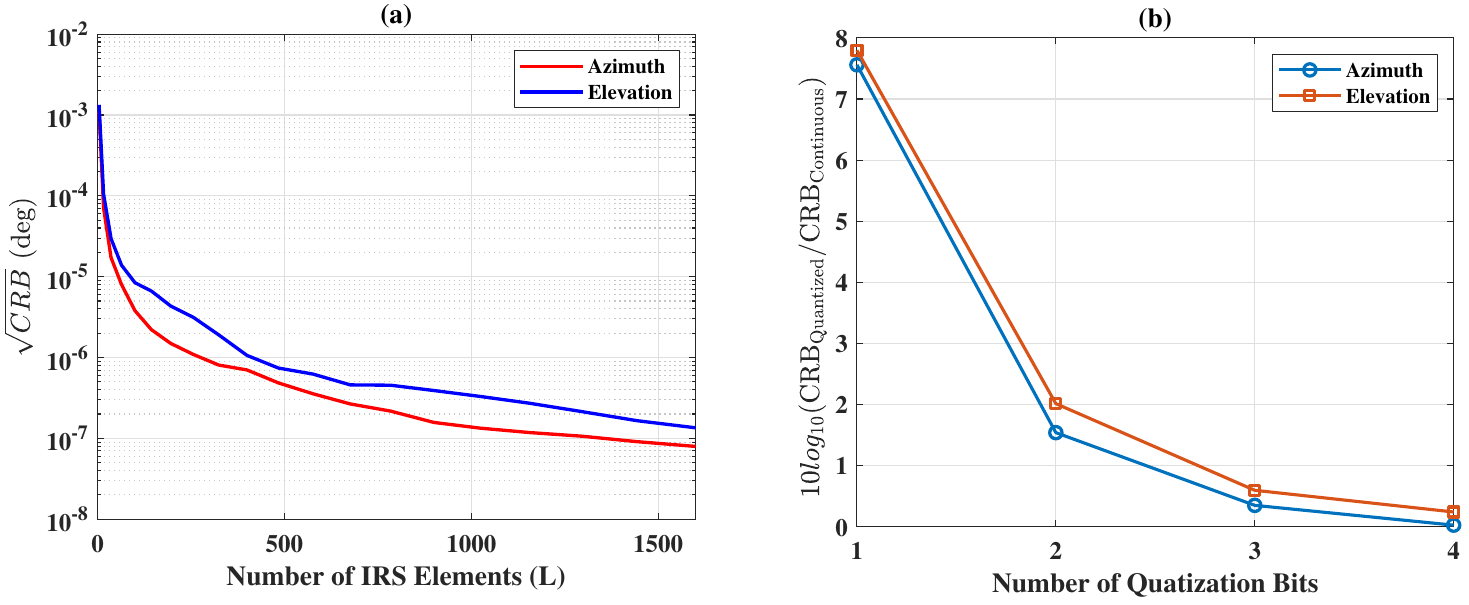}
    \caption{(a) CRB as a function of the number of IRS elements. and (b) Relative CRB versus Quantization Levels of IRS.}
    \label{fig:rmse_irs_n_quantiz}
\end{figure*}

Following our study of the target, we will now examine the sensitivity of parameter estimation to various factors related to IRS. First, to determine the optimal placement of the IRS panel, we fix the radar position and FOV region. Keeping the round-trip delay constant, we treat the radar position and the FOV point as the foci of an ellipse, place the IRS panel along the ellipse, and analyse the resulting parameter-estimation performance. As shown in Fig. ~\ref{fig:rmse_irs_placement}(a), starting from the original IRS placement considered, we move the IRS along the defined elliptical path towards (negative angles) and away (positive angles) from the radar. Fig. ~\ref{fig:rmse_irs_placement}(b) illustrates the effect of placing the IRS panel at different points along the elliptical path. When the IRS moves toward the FOV region, the CRB increases as the path loss in the IRS-radar channel becomes substantial due to the greater distance between the IRS and the radar. When the IRS panel moves toward the radar, the CRB decreases, indicating improved performance. This effect can be attributed to the effective path loss over the entire signal round trip. The effective path loss varies as the product of the path loss of each round-trip segment changes, while the total path delay remains constant at every instance. However, the lowest CRB occurs when the IRS panel is positioned behind the radar. This placement is not optimal for the configuration, as it would largely mimic the monostatic path that is not suitable for stealth target detection. Therefore, to improve detection and estimation performance for stealth-target detection, the IRS panel should be placed at a suitable distance in the radar's left/right frontal region.

Figure~\ref{fig:rmse_irs_n_quantiz}(a) illustrates the dependence of azimuth and elevation CRB on the number of IRS elements, $L=L_h L_v$. In this study, as the number of IRS elements increases, we maintain a square IRS panel with $L_h=L_v\in[2,40]$. As $L$ increases from $4$ to $1600$, the azimuth angle and elevation angle CRB decrease, proving the role of aperture gain in IRS-assisted sensing. Larger IRS apertures enhance the effective array gain and enable more accurate phase synthesis, thereby strengthening the reflected signals and improving angular resolution. The CRB reduction is most pronounced in the range $4 \leq L \leq 400$, after which the curves tend to saturate, indicating diminishing returns with further aperture enlargement for the given set of system parameters. The CRB of azimuth is lower than that of elevation, following the similar trend observed in Figs.~\ref{fig:rmse_snr_ss}(a) and ~\ref{fig:rmse_snr_ss}(b). Overall, increasing the number of IRS elements is particularly beneficial in the low- to mid-aperture regime, whereas for larger apertures the design must balance incremental performance gains against the FOV and sidelobe attenuation.

Once we have determined the optimal number of elements and the position of the IRS, we examine the design of IRS weights in a real-time hardware scenario. Specifically, we focus on how quantized weight implementation on the IRS panel affects parameter estimation performance. We design the IRS phase weights using the method described in Section~\ref{irsweightdesign} and then quantize them at different levels (1-bit, 2-bit, 3-bit, and 4-bit). We calculate the CRB for each quantization level and compare with the CRB obtained for continuous weights (the ideal case) as a ratio. This ratio provides insight into the relative estimation error associated with quantised weights and is plotted across the different quantization levels. The results are shown in Fig.~\ref{fig:rmse_irs_n_quantiz}(b). The figure shows that the CRB decreases as the number of quantisation bits increases, attributing to the phase quantisation levels approaching towards the ideal phase weights.

Furthermore, for the design of IRS weights as illustrated in (\ref{WeightsSoln}), it is to be noted that the beamwidth decreases with increasing aperture length, thereby affecting the FOV. We remark that the design of IRS weights using well-constructed optimization problem takes the center stage in the performance of the IRS-static radar configuration for the desired FOV which would be potential study for future work.


\section{Conclusion}\label{conclusion}
 In this work, a three-hop IRS-assisted radar configuration, termed an IRS-static radar, is investigated. It is shown that this configuration emulates a bi-static radar while alleviating the need for time synchronization, as the IRS panel redirects the scattered signal back to the monostatic radar. A signal model for this configuration is formulated and expressed compactly to derive the CRB for the target parameters. The dependence of the CRB on key system parameterssuch as the signal-to-noise ratio (SNR), number of snapshots, IRS beamforming weights, IRS placement, and number of reflecting elements is investigated through extensive simulations. Motivated by real-world deployment considerations, particular emphasis is placed on practical design aspects, including optimal IRS placement and realizable IRS phase-weight configurations. The variation of the CRB with target movement in both azimuth and elevation is analyzed to characterize the system’s achievable field of view and sensing coverage. The results show that the performance of the proposed IRS-static radar depends strongly on the design and placement of the IRS. Performance is also affected by the quality of the channel between the IRS and the radar, making system geometry and IRS configuration important design considerations. This study provides useful design guidelines for practical IRS-static radar system and serves as a benchmark for developing and evaluating parameter-estimation algorithms for this architecture. However, developing such parameter-estimation techniques suited to this configuration is a non-trivial extension of existing methods and can be pursued as part of continuation of this research work.

\bibliographystyle{IEEEtran}
\bibliography{ref}

\end{document}